\documentclass[12pt]{article}
\usepackage{graphicx}
\def\tt{$t\bar{t}$}
\def\bb{$b\bar{b}$}

\newcommand\pubnumber{FERMILAB-CONF-14-410-E}
\newcommand\pubdate{October 23, 2014}

\def\fmph{Faculty of Mathematics, Physics and Informatics, \\
Comenius University, \\ Mlynska Dolina F1, 842 48 Bratislava, Slovakia}
\def\support{\footnote{Work supported by the Ministry of Education, Science, Research and Sport of the Slovak Republic}}

\def\Title#1{\begin{center} {\Large #1 } \end{center}}
\def\Author#1{\begin{center}{ \sc #1} \end{center}}
\def\Address#1{\begin{center}{ \it #1} \end{center}}

\newcommand\pubblock{\rightline{\begin{tabular}{l} \pubnumber\\
         \pubdate  \end{tabular}}}
\newenvironment{Abstract}{\begin{quotation}  }{\end{quotation}}
\newenvironment{Presented}{\begin{quotation} \begin{center} 
             PRESENTED AT\end{center}\bigskip 
      \begin{center}\begin{large}}{\end{large}\end{center} \end{quotation}}
\def\Acknowledgements{\bigskip  \bigskip \begin{center} \begin{large}
             \bf ACKNOWLEDGEMENTS \end{large}\end{center}}

\def\beq{\begin{equation}}
\def\eeq#1{\label{#1}\end{equation}}
\def\eeqn{\end{equation}}

\def\beqa{\begin{eqnarray}}
\def\eeqa#1{\label{#1}\end{eqnarray}}
\def\eeqan{\end{eqnarray}}

\let\bar=\overbar

\def\etal{{\it et al.}}

\def\Dslash{\not{\hbox{\kern-4pt $D$}}}
\def\dslash{\not{\hbox{\kern-2pt $\del$}}}

\def\msb{{\bar{\ssstyle M \kern -1pt S}}}

\def\Journal#1#2#3#4{{#1} {\bf #2}, #3 (#4)}

\def\PRL{Phys. Rev. Lett.}
\def\PRD{Phys. Rev. D}

\begin{document}
\begin{titlepage}
\pubblock

\vfill
\Title{Asymmetries at the Tevatron}
\vfill
\Author{ Pavol Barto\v{s} on behalf of CDF and D0 Collaborations\support}
\Address{\fmph}
\vfill
\begin{Abstract}
In this report, we summarize the latest results of the top-quark pair production asymmetry and present the new result of bottom-quark pair production asymmetry. By looking at the results obtained by the CDF experiment, one can see a discrepancy in both \tt{} inclusive and lepton-based measurements. The D0 results of the \tt{} production asymmetry are compatible with the standard-model predictions as well as with the CDF results. The CDF measurement of \bb{} production asymmetry presents consistency with both zero and with the standard-model predictions.
\end{Abstract}
\vfill
\begin{Presented}
the 8th International Workshop on the CKM Unitarity Triangle (CKM 2014),\\ Vienna, Austria, September 8-12, 2014
\end{Presented}
\vfill
\end{titlepage}
\def\thefootnote{\fnsymbol{footnote}}
\setcounter{footnote}{0}

\section{Introduction}

The top quark is the heaviest fundamental particle with unique properties. Due to the very short lifetime ($\sim 10^{-25}$\ s), the top quark decays before hadronization and can be studied using its decay products. If a deviation of the measured properties from the standard-model (SM) predictions is seen, it could be a sign of new physics. At the Tevatron $p\bar{p}$ collider, top quarks are produced mainly in pairs through strong force quark-antiquark annihilation ($\sim 85$\%) and gluon-gluon fusion ($\sim 15$\%) processes. According to the SM, the top quark decays into the $W$ boson and bottom ($b$) quark in almost 100\% of the cases. The final state of top-quark-pair production contains two $b$-quark jets and two $W$ bosons, which decay leptonically (to $l\nu_{l}$, where in our case $l = e,\mu$) or hadronically (into quarks). The $t\bar{t}$ events can then be classified into three categories: the $dilepton$ or $all$-$jets$ events, where both $W$ bosons decay leptonically or hadronically, respectively; and the $lepton$$+$$jets$ events, where one of the $W$ bosons decays leptonically while the other one decays hadronically.

The bottom quark is the second heaviest fundamental particle with mass $\sim$$40$-times lower than the mass of the top quark. The bottom quarks can be studied directly using the jets produced during hadronization. At a hadron collider \bb{} pairs are mainly produced by the gluon-gluon fusion process.

\section{\tt{} production asymmetries}\label{sec:asym}
The $t\bar{t}$ pairs are produced through strong force quark-antiquark annihilation or gluon-gluon fusion. In the leading order (LO), the SM does not predict any asymmetry in either of the processes. However, at NLO, the SM predicts an asymmetry coming from interference of the amplitudes of Born and box diagrams and interference of the initial and final state gluon radiation in quark-antiquark annihilation processes. The $t\bar{t}$ production via gluon-gluon fusion remains symmetric also at higher orders. Another contribution to asymmetry is due to interference of quark-gluon scattering processes or due to the electroweak interactions, but the expected contribution is small.

\subsection{\tt{} forward-backward asymmetry}
After the kinematic reconstruction of final $t\bar{t}$ state, one can define the forward-backward asymmetry using rapidity difference, $\Delta y = y_t - y_{\bar{t}}$:
\begin{equation}
A_{FB} = \frac{N(\Delta y > 0) - N(\Delta y <0)}{N(\Delta y > 0) + N(\Delta y <0)}
\label{eq:afb}
\end{equation}
\noindent
where $y_t$ ($y_{\bar{t}}$) corresponds to rapidity of top (antitop) quark.

\begin{figure}[h!]
\begin{minipage}{0.48\linewidth}
\centering{\includegraphics[width=0.95\linewidth]{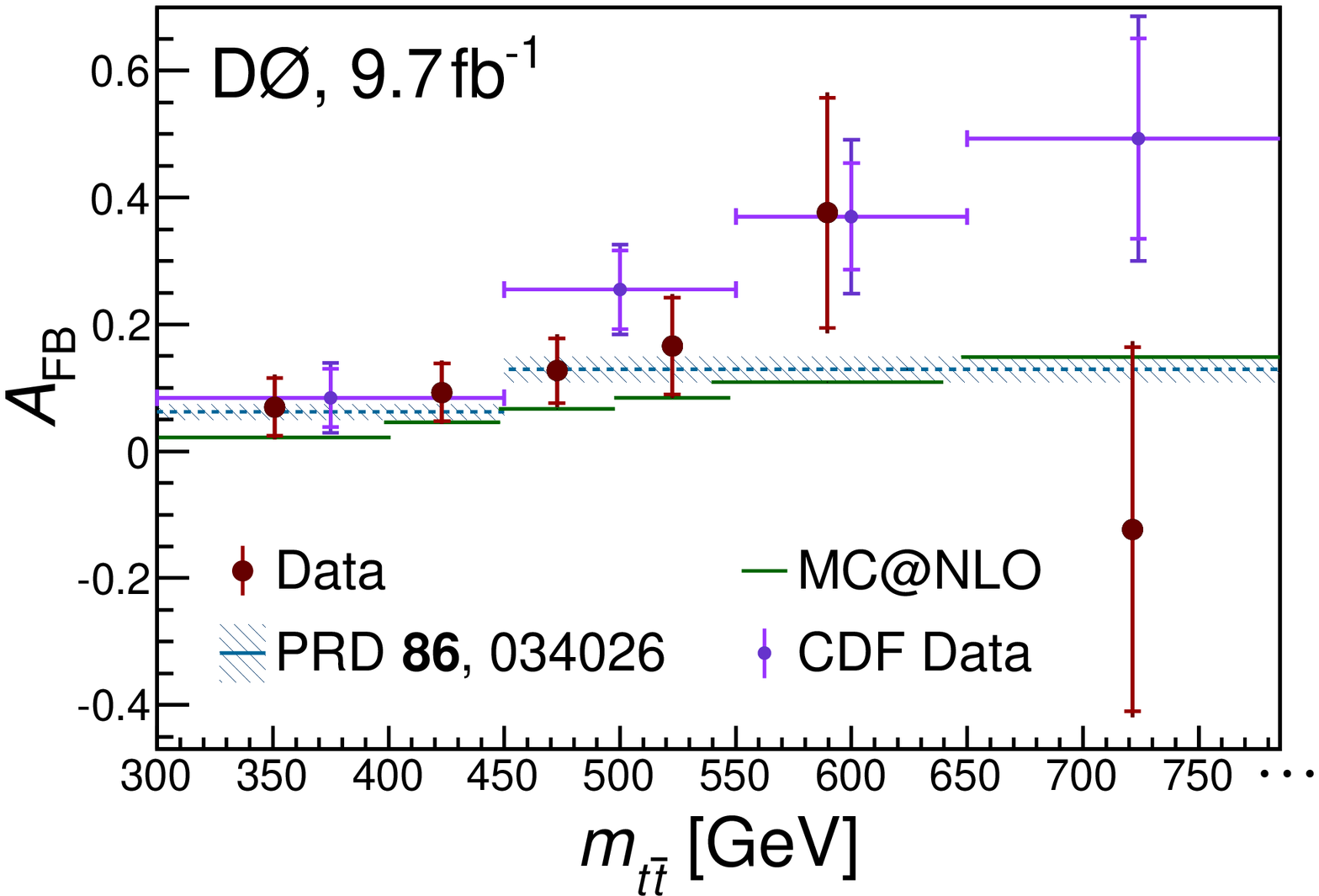}}
\caption{Forward-backward asymmetry as a function of top-quark pair mass at the parton level. D0 and CDF results are shown in black and purple points, respectively.}
\label{fig:afb}
\end{minipage}
\hfill
\begin{minipage}{0.48\linewidth}
\centering{\includegraphics[width=0.95\linewidth]{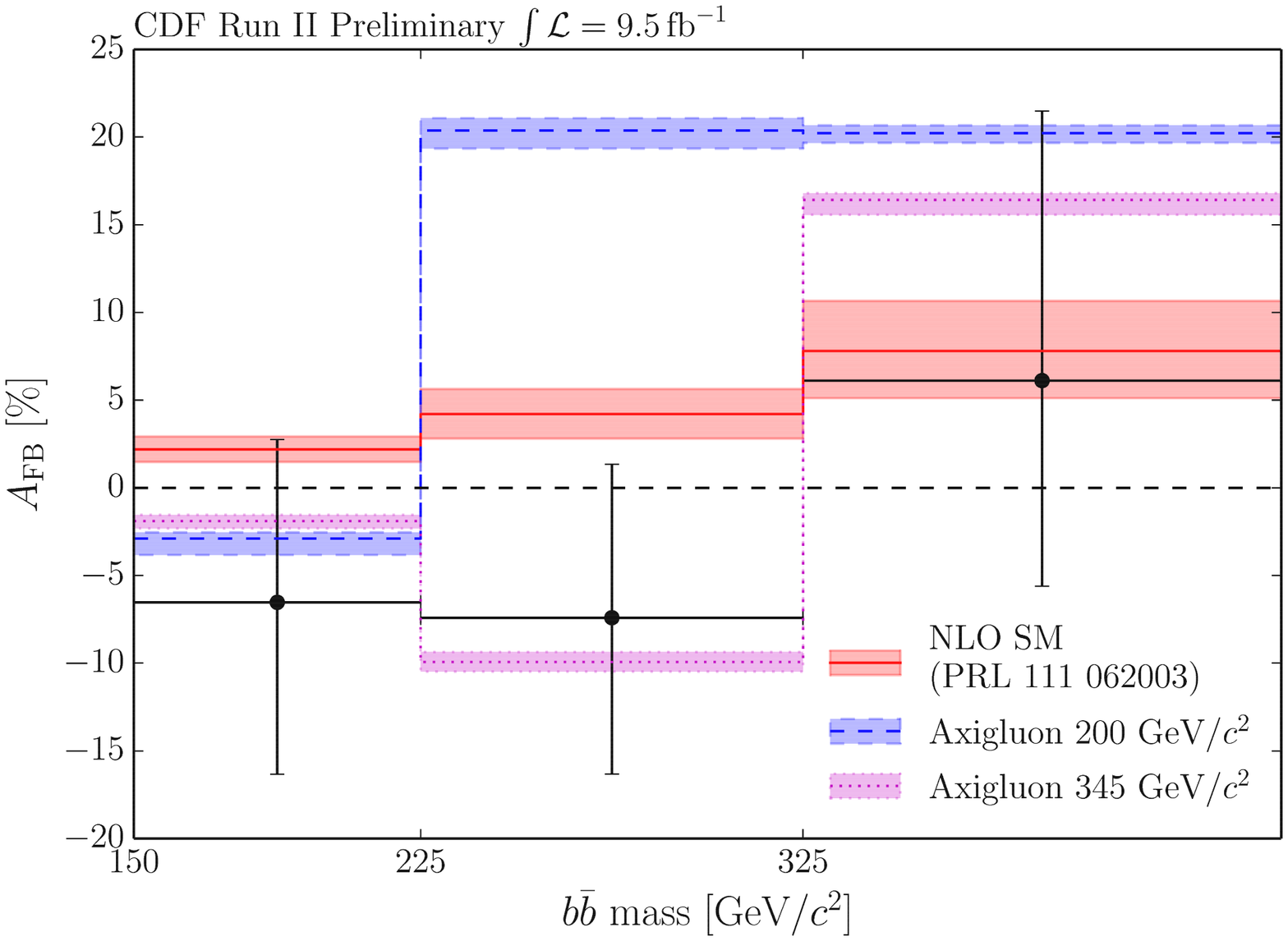}}
\caption{The measured particle-jet level \bb{} asymmetry in each mass bin. The error bars represent the 68\% credible intervals.}
\label{fig:bbafb}
\end{minipage}
\hfill
\end{figure}

Both CDF and D0 measure the asymmetry in the $lepton+jet$ channel using full data sets. After the full kinematic reconstruction of $t\bar{t}$ final state, the correction to the parton level is done using regularized 2D unfolding. The asymmetry measured by the CDF experiment is $A_{FB} = 0.164 \pm 0.039$(stat) $\pm 0.026$(syst)~\cite{CDFafb}, while the value measured by the D0 experiment is $A_{FB} = 0.106 \pm 0.027$(stat) $\pm 0.013$(syst)~\cite{D0afb}. 

Both experiments express the inclusive asymmetry as a function of top-quark pair mass, $m_{t\bar{t}}$, and modulus of rapidity difference, $|\Delta y|$. In addition to that, CDF presents also dependence of $A_{FB}$ on the transverse momentum of the $t\bar{t}$ system, $p_{T}(t\bar{t})$. The asymmetry at CDF is found to have approximately linear dependence on both $|\Delta y|$ and $m_{t\bar{t}}$, as expected for the NLO charge asymmetry, although with larger slopes than are expected in the NLO prediction. The probabilities to observe the measured values or larger for the detector-level dependencies are 2.8$\sigma$ and 2.4 $\sigma$ for $|\Delta y|$ and $m_{t\bar{t}}$, respectively. The D0 measurements of the $A_{FB}$ dependences on $m_{t\bar{t}}$ and $|\Delta y|$ show compatibility of the data with the SM predictions as well as with the CDF results (see Fig.~\ref{fig:afb}).

The solution of the puzzling situation could come from the theoretical calculations at NNLO. According to the talk of A. D. Mitov~\cite{xsecpred}, the D0 results are in good agreement with the predictions at NNLO, while the CDF results are a bit larger (but less than 1.5$\sigma$). 

\subsection{Lepton-based \tt{} asymmetry}

To measure the lepton-based asymmetry, there is no need to reconstruct the $t\bar{t}$ final state. The advantage is also a good lepton charge determination and high precision of measurement of the lepton direction. One can define single-lepton asymmetry, $A_{FB}^{l}$, using the lepton charge multiplied by its rapidity ($qy_{l}$); or dilepton asymmetry, $A_{FB}^{\Delta\eta}$, using the difference of pseudo-rapidities of positive and negative leptons ($\Delta\eta = \eta_{l^{+}} - \eta_{l^{-}}$) in dilepton channel. 

Both experiments measure the single-lepton asymmetry in $lepton+jets$~\cite{CDFljafb,D0ljafb} and $dilepton$~\cite{CDFdilafb,D0dilafb} channels and combine the results using the BLUE method. CDF obtains $A_{FB}^{l}=0.090^{+0.028}_{-0.026}$~\cite{CDFdilafb}, while D0 measures the value of $A_{FB}^{l}=0.042 \pm 0.020$(stat)$\pm 0.014$(syst)~\cite{D0ljafb}. Comparing the results with the SM prediction of $A_{FB}^{l}=0.038 \pm 0.003$, one can say that CDF sees a $2\sigma$ excess, while the D0 result is compatible with expectations. 

In the $dilepton$ channel, CDF and D0 measure a dilepton asymmetry of $A_{FB}^{\Delta\eta}=0.072 \pm 0.081$~\cite{CDFdilafb} and $A_{FB}^{\Delta\eta}=0.123 \pm 0.054$(stat)$\pm 0.015$(syst)~\cite{D0dilafb}, respectively. Both results are compatible with the SM prediction of $0.048 \pm 0.003$. Furthermore, D0 presents a ratio of single-lepton and dilepton asymmetries in the $dilepton$ channel. There is a discrepancy between the measured value of the ratio of $A_{FB}^{l}/A_{FB}^{\Delta\eta}=0.36\pm0.20$ and the SM prediction of $0.79\pm0.10$.

\section{\bb{} forward-backward asymmetry}
The \bb{} production at a hadron collider is almost exclusively a QCD process. To be able to measure a forward-backward asymmetry one has to select a kinematic region where the $q\bar{q}$ initial state is significantly enhanced over the symmetric gluon-gluon fusion processes. There are more quarks at large Bjorken $x$; therefore, $gg\rightarrow b\bar{b}$ processes can be suppressed by requiring a large mass $m_{b\bar{b}}$ of the \bb{} pair.  

The measurement uses dijet events. Two jets are required to be tagged as jets initiated by a $b$- or $\bar{b}$-quark. The \bb{} pair mass, $m_{b\bar{b}}$, has to be above $150\:$GeV.

The \bb{} forward-backward asymmetry can be defined by Eq.~(\ref{eq:afb}) while the rapidity difference, $\Delta y$, is now defined as the difference of the $b$-quark ($y_b$) and $\bar{b}$-quark ($y_{\bar{b}}$) rapidities. To distinguish if the $b$-jet was initiated by a $b$- or $\bar{b}$-quark, the momentum-weighted average of the charges of the tracks associated with the $b$-jet ($b$-jet charge) is used. The forward/backward assignment of the event is done by computing the charge difference of the two $b$-jets.


To extract the result at the particle-jet (hadron-jet) level, the Bayesian technique is used. A formula relates the background asymmetry, charge confusion rate, \bb{} fraction in data, mass smearing, and signal asymmetry. Fig.~\ref{fig:bbafb} shows the measured particle-jet level asymmetry in three $m_{b\bar{b}}$ bins. The result~\cite{bbres} is consistent with zero, the SM prediction~\cite{bbpred} and with the $345\:$GeV axigluon model. The $200\:$GeV axigluon model is inconsistent with the measurement at more than $95\%$. 

\section{Conclusions}
The measurements presented here are mostly in agreement with the standard-model predictions. The CDF sees higher production asymmetry in both \tt{} inclusive and lepton-based measurements. The D0 data are compatible with standard-model predictions and also with the CDF results. Including NNLO correction in the theoretical predictions seems to be promising in solving this puzzling situation. The first measurement of the bottom forward-backward asymmetry at high \bb{} mass presents consistency with both zero and with the standard-model predictions. 

\Acknowledgements
It is a pleasure to thank the CDF and D0 collaborators for their well-done work, the top-group conveners for their help and the organizers of the CKM 2014 for a very interesting conference. This work was supported by the Ministry of Education, Science, Research and Sport of the Slovak Republic.


\begin{thebibliography}{99}

\bibitem{CDFafb} T. Aaltonen \etal, (CDF Collaboration), \Journal{\PRD}{87}{092002}{2013}.
\bibitem{D0afb} V.M. Abazov \etal, (D0 Collaboration), Fermilab-Pub-14-116-E, arXiv:1405.0421 [hep-ex], accept. by PRD.
\bibitem{xsecpred} A. D. Mitov, Top quark pair cross section, talk at CKM 2014, Wien, September 8-12, 2014 (C14-09-08).
\bibitem{CDFljafb} T. Aaltonen \etal, (CDF Collaboration), \Journal{\PRD}{88}{072003}{2013}.
\bibitem{D0ljafb} V.M. Abazov \etal, (D0 Collaboration), \Journal{\PRD}{90}{072001}{2014}.
\bibitem{CDFdilafb} T. Aaltonen \etal, (CDF Collaboration), \Journal{\PRL}{113}{042001}{2014}.
\bibitem{D0dilafb} V.M. Abazov \etal, (D0 Collaboration), \Journal{\PRD}{88}{112002}{2013}.
\bibitem{bbres} T. Aaltonen \etal, (CDF Collaboration), CDF Conference Note 11092.
\bibitem{bbpred} B. Grinstein, Ch. W. Murphy, \Journal{\PRL}{111}{062003}{2013}.


\end{thebibliography}
\end{document}